\documentclass[12pt,a4paper, compsoc]{IEEEtran}
\usepackage{fancyhdr}
\usepackage{footnote}
\usepackage{cite}
\usepackage{lscape}
\usepackage[dvips]{graphicx}
\usepackage{makecell}
\usepackage{amsfonts}
\usepackage{caption}
\usepackage{subcaption}
\usepackage{multirow}
\usepackage[table,xcdraw]{xcolor}
\usepackage{booktabs}
\usepackage{tabularx}
\newcolumntype{L}[1]{>{\raggedright\arraybackslash}p{#1}}
\newcolumntype{C}[1]{>{\centering\arraybackslash}p{#1}}
\newcolumntype{R}[1]{>{\raggedleft\arraybackslash}p{#1}}
\usepackage{listings}
\usepackage[hyphens]{url}
\usepackage[colorinlistoftodos]{todonotes}
\usepackage{supertabular}
\usepackage{ragged2e}
\usepackage{ifsym}
\usepackage{marvosym}
\usepackage{textcomp}
\usepackage{makecell}
\usepackage{tabu}
\usepackage{amsmath}
\usepackage{longtable}
\usepackage{supertabular}
\usepackage{csvsimple}

\usepackage{hyperref}
\hypersetup{
    pdftitle={Technology Mapping Using WebAI: The Case of 3D Printing},
    colorlinks=true,
    linkcolor=blue,
    filecolor=blue,
    urlcolor=blue,
    citecolor=blue
    }

\title{Technology Mapping Using WebAI:\\ The Case of 3D Printing}

\author{
    \IEEEauthorblockN{
        Julian~Schwierzy$^1$, 
        Robert~Dehghan$^{4,6}$, 
        Sebastian~Schmidt$^{5,6}$,
        Elisa~Rodepeter$^1$,
        Andreas~Stömmer$^1$, 
        Kaan~Uctum$^1$, 
        Jan~Kinne$^{3,6}$, 
        David~Lenz$^{6,7}$, Hanna~Hottenrott$^{1,2,3}$ 
    }
}

\IEEEtitleabstractindextext{
    \small
        $^1$Technical University of Munich,
        $^2$Munich Data Science Institute,
        $^3$Leibniz Centre for European Economic Research (ZEW),
        $^4$University of Mannheim,
        $^5$University of Salzburg,
        $^6$ISTARI.AI,
        $^7$Justus Liebig University Giessen; 
        \Letter{ julian.schwierzy@tum.de} \\
	
 \begin{abstract}
    \justifying
    The diffusion of new technologies is crucial for the realization of social and economic returns to innovation. Tracking and mapping technology diffusion is, however, typically limited by the extent to which we can observe technology adoption. This study uses website texts to train a multilingual language model ensemble to map technology diffusion for the case of 3D printing. The study identifies relevant actors and their roles in the diffusion process. The results show that besides manufacturers, service provider, retailers, and information providers play an important role. The geographic distribution of adoption intensity suggests that regional 3D-printing intensity is driven by experienced lead users and the presence of technical universities. The overall adoption intensity varies by sector and firm size. These patterns indicate that the approach of using webAI provides a useful and novel tool for technology mapping which adds to existing measures based on patents or survey data.
    \textbf{Data:} An interactive visualisation of our data can be found under \url{https://stories.istari.ai/3D/en/}.
  \end{abstract}

}

\begin{document}

\maketitle

\thispagestyle{fancy}
\pagestyle{fancy}

\begingroup\renewcommand\thefootnote{\textsection}

\section{Introduction}
\renewcommand{\thefootnote}{\arabic{footnote}}
\IEEEPARstart{N}{ew} technologies are a key driver of sustainable economic development and a crucial contributor to the substantial rise in living standards the world has seen since the first industrial revolution \cite{AghionAntoninBunel+2021}. Technological innovation is also important for the competitiveness of companies and their performance \cite{Audretsch1996,Crepon1998,Aghion2009}. Policy-makers and practitioners are therefore interested in understanding the emergence of new technologies. Yet, it is not only the invention that matters, but also the adoption and diffusion of innovation \cite{Hall2006}.
Tracking the adoption of new technologies is, however, not straightforward. Researchers typically rely on information drawn from patent applications to measure, understand, and localize inventive activity \cite{MOSER2013,Pose-Rodriguez2020}. While such data is extremely useful, it applies only to patentable inventions, whereas innovation and diffusion in other areas may not be captured \cite{Hall2020}.\\ 
Especially, for an interdisciplinary technology such as 3D printing, the identification of relevant patents is a challenge, as these can be filed in very different patent technology classes \cite{Pose-Rodriguez2020}. Moreover, understanding the spread of new technologies require information on different actors relevant to diffusion processes. In addition to the manufacturers and users of the technology, others also contribute to the diffusion by simply providing information, thus helping to raise customer awareness and reduce the adoption resistance of potential users and intermediaries. Alternatively to patent information, company surveys on new products, processes, and services are useful for collecting information that augments patent-based measures. Yet, surveys are typically restricted to smaller samples and may potentially overlook significant to the diffusion processes in certain areas \cite{Salazar2004}. 
In this study, we therefore propose a novel methodology for mapping technology diffusion in areas where patent data or company surveys may not be sufficient. 
We follow the approach recently developed by Kinne et al. \cite{Kinne2021} in the context of analyzing the reactions of German companies to the COVID-19 pandemic. The webAI method builds on advances in research on natural language processing to measure company activity (e.g. innovation) using web mining and deep learning \cite{Kinne2021}. 
In particular, we analyze text information retrieved from company websites to map the diffusion of 3D printing technologies while distinguishing between different contributors. For data collection, a scraper screens for topic-specific keywords in text paragraphs on company websites. By manually labeling a subset of the identified text paragraphs, we build a training data set based on the context in which the keywords appear. A model ensemble is then trained on the labeled data and subsequently applied to the companies with web domains in Germany listed in the ORBIS firm database (approx. 1.3~million companies). Firms use their website to promote their products and services, but also to inform the public about important events. Websites can therefore be used to read out the activities of companies, which is facilitated by the fact that, depending on their size and industry, almost every relevant company has its own website \cite{Kinne2020}.

The model yields predictions for the degree of engagement at a highly granular regional level as well as a classification for the type of involvement of all included companies in 3D printing activities. We differentiate the four types: \textit{Manufacturers}, \textit{Service}, \textit{Retailer} and  \textit{Information}. Based on this fine-grained data, we can map the diffusion of 3D printing in Germany and study the adoption intensity across sectors and regions. 
We find several adoption hotspots and show that the local adoption intensity seems to be related to two factors: the closeness to technical universities and the presence of traditional manufacturing companies. Both appear to contribute to a more intense use of 3D printing as a new technology, possibly through different channels including access to human capital, opportunities for applications, and customers. These results are in line with earlier research that stressed the importance of the co-location of science and industry for spurring innovation especially in emerging technologies \cite{Hausman2021,Cohen2002}.     

\section{Background on 3D Printing}
3D printing, often used synonymously with the term additive manufacturing, is an umbrella term for a group of production technologies\footnote{While additive manufacturing is generally defined as a process of combining materials to create objects layer by layer, 3D printing is specifically the fabrication of objects using a print head, nozzle or other printer technologies \cite{ISO/ASTM2018}.}. 
A key characteristic is the use of digital designs for creating a physical product layer by layer \cite{gebhardt2016additive}. Compared to conventional manufacturing technologies such as drilling, milling, and injection moulding, 3D printing provides substantial advantages in terms of sustainability \cite{ghobadian2020examining}, flexibility \cite{Eyers2018}, and the freedom to design complex geometries \cite{gebhardt2016additive}. Thus, it allows rapid idea iterations to create new goods \cite{weller2015economic}, decentralized production \cite{ben2017decentralization}, and it affects competition \cite{schwierzy2021digitalisation}.\\
The development of 3D printing technology started in the 1980s, when the main applications included the production of prototypes and spare parts. More recently, 3D printing has also become more important for the production of end-products \cite{su2018history}. Its increasing relevance is also reflected in the continuous growth of the 3D printing market including revenues from 3D printing systems, materials, software, and services. A compound annual growth rate of roughly 25\% since 2014 led to a global market size of 3.7~billion EUR in 2014, 8.9~billion EUR in 2018, and is predicted to reach approximately 30.2~billion EUR in 2024 \cite{Pose-Rodriguez2020}. Overall, actors relevant to the adoption and diffusion of 3D printing technology comprise manufacturers of the equipment (machines, materials, and software) on the supply side and technology adopters on the demand side. In addition, information providers such as consulting companies, blogs, and associations contribute to the diffusion of knowledge about 3D printing. Despite these insights, we still know little about the diffusion of 3D printing technology across sectors and geographic regions as well as the role of different actors in this process.  

\section{Methodology}
The main objective of this study is to map the diffusion of 3D printing technology using information that is readily available and by following a transparent and reproducible approach. For this purpose, we make use of company websites and general company-level information obtained from the ORBIS database. The web analysis of this study is conducted using webAI, a cloud-based and artificial intelligence web analysis tool set developed by ISTARI.AI. With the help of this tool we aim at identifying and clustering firms according to their role in the innovation and diffusion process of the 3D printing technology. Our webAI based approach comprises several steps, which we describe in the following. 

\subsection{Identification of Keywords}
In order to determine 3D printing technology engaged companies, we identify a list of keywords related to 3D printing. A uniform standardized terminology of 3D printing technologies does not exist. However, companies typically use the technical terms from industrial standards, in this case from ASTM~52900 and VDI~3405. These lists of terms contain the nomenclature of different technologies and their applications. Due to the constant and rapid change of 3D printing technology, we augment these keywords with terms identified in current research papers and publications by a consulting company. A complete list of all keywords used for the search is attached in the appendix (see Table~\ref{table:clusterDescriptions}).   

\subsection{Identification of Types}
Our initial data contains 1.3~million economically active companies in Germany listed in the ORBIS database. This data set consists of a corporate URL as well as company characteristics, e.g. number of employees and incorporation date. The definition of types was performed in several feedback loops. A first data set was generated in May 2021 by searching the corporate websites for the selected keywords. This led to 103,000~data points, each containing the URL of the website where the keyword was found, the keyword itself, as well as the paragraph including the respective keyword. If a URL included multiple keywords, multiple data points were created. 
In the first step, a data subset of 750~data points was selected randomly and divided evenly among five different researchers, who then made an initial proposal for the type. For this first proposal, both the paragraphs and the overall website itself were analyzed. Since the resulting classification proposals strongly resembled each other, we agreed on seven types in a first feedback round: \textit{Manufacturer}, \textit{Service}, \textit{3D Printing for own Products}, \textit{Consulting \& Education}, \textit{Retail}, \textit{Information}, and \textit{Others}. 
Those defined types were then used to manually label a training data set. For this purpose, 3,000~data points were randomly selected from the initial 103,000~data points. Again, the labels were assigned on the basis of a keyword, a corresponding paragraph, and the URL. To avoid any bias due to prior knowledge about the companies, this data subset differed from the first data subset used to define the types. The decisive factor for the respective label should be the given paragraph. The labelling of this data set was again carried out by multiple researchers. We used the labeled data set to train the model. In several feedback rounds, the prediction results of the model were assessed using a test data set to gauge the prediction performance. It became clear that certain highly similar categories and their associated keywords were difficult to distinguish for the model. This led to the adjustments of keywords and labels. Eleven keywords that were not precise enough for identifying 3D printing technologies, i.e. they often appeared in a different context, were removed. See Appendix Table~\ref{table:Appendix_3d_printing_keywords} for a list of initial and final keywords. Data points stemming from keywords that only appeared in menus, signatures, or similar non-content parts of websites were also removed from the data set. This resulted in a decrease of the data set from 103,000 to 32,000~data points. Further, the models showed a low accuracy in distinguishing between the labels \textit{Service}, \textit{Consulting \& Education} and \textit{3D Printing for Own Products}. Therefore, we decided to combine those labels. We dropped the label \textit{Others} due to a low number of allocations, resulting in four final types: \textit{Manufacturer}, \textit{Service}, \textit{Retail}, and \textit{Information}.
A detailed description of those types can be found in Table \ref{table:clusterDescriptions}.

\begin{table*}
\centering
\caption{Labels for Website Classification}
\label{table:clusterDescriptions}
\begin{tabular}{L{9.5cm} L{4.5cm} L{3cm}}
 \textbf{Definition} & \textbf{Initial Classification}  & \textbf{Final Classification} \\
 \hline 
 Manufacturer of 3D printers or equipment (e.g. software, material) & Manufacturer & Manufacturer \\ 
 Offering personalized 3D printing services & Service  & Service\\ 
 Using 3D printing technologies in own production & 3D Printing for own Products & Service \\ 
 Offering consulting, training etc. for firms to adopt 3D printing & Consulting \& Education &  Service\\ 
 Offering 3D printers, material, spare parts, etc. & Retail & Retail \\ 
 Providing information about 3D printing technologies & Information & Information\\ 
Miscellaneous purposes & Others & $-$ \\ 
 \hline
\end{tabular}
\end{table*}

\subsection{Model Creation}
Based on the manually labeled data set, we trained a webAI model to automatically label the context of 3D printing related web text content according to the four pre-defined types. The final classification system was a model ensemble consisting of 10~single models. Each of the 10~constituent models received different data during training, thus learning distinct patterns from the data. Consequently, each model became an expert in differentiating the defined types. However, each model used a slightly different reasoning. As each model cast a vote during inference, the type with the most votes became the final prediction. Ensemble models greatly increase the reliability of the results, as single models can easily over- or under-fit the training data, leading to unwanted edge cases and overall sub-par performance. This procedure also allows to calculate confidence scores for the predictions, i.e. the more models favor the same outcome, the higher the confidence in the decision. We further optimized each constituent models architecture through an extensive neural architecture search\cite{jin2019auto}.
The constituent models received semantic vectors as inputs. The semantic vectors resulted from encoding the relevant paragraphs using pre-trained sentence transformers \cite{reimers2019sentencetransformers}. The advantage of pre-trained models is that they already possess fundamental language understanding \cite{malte2019evolution}. Thus, only comparatively little training data and minor adaptions are necessary to cope with new use cases. Along with the relevant text parts, we further encoded several company and website meta-information and concatenated them into a single high-dimensional semantic vector. In our specific case, a 1,920~dimensional semantic vector represented the relevant information about each company. 


Since a company's website usually consists of several sub-webpages and each sub-webpage is labeled individually, a company can be labeled with multiple types, e.g. \textit{Information} and \textit{Retail}. To classify the entire company, we introduced a hierarchy system. The final classification of a company corresponded to the highest ranking label assigned to that company. The chosen hierarchy is shown in Table 1. For example, a company that provides 3D printing services but also manufactures 3D printers would be assigned the final label \textit{Manufacturer}.

\section{Results}
We present our findings in two steps. First, we describe the characteristics of the firms involved in 3D printing and their respective types. Second, we study the geographical distribution of 3D printing companies and relate the occurrence of 3D printing hotspots to the location of relevant manufacturing clusters and universities.

\subsection{Company characteristics}
When differentiating between roles that the companies involved in 3D printing take in the adoption and diffusion of the technology, we find that the vast majority of companies (total: 6,336) act as service providers (71.9\%). 19.7\% of companies provide information as the main activity. Manufacturers comprise 8.0\% of companies and retail 0.4\%.

Dividing the engaged companies into six age classes shows that both very young as well as relatively old firms are active in the area of manufacturing and service provision. In retail, mainly medium-old companies are most active. Interestingly, information providers are relatively established in terms of age suggesting that experience and credibility may matter here (see Figure \ref{fig:3D_Companies_Age}). 

\begin{figure}
    \centering
    \includegraphics[scale=0.43]{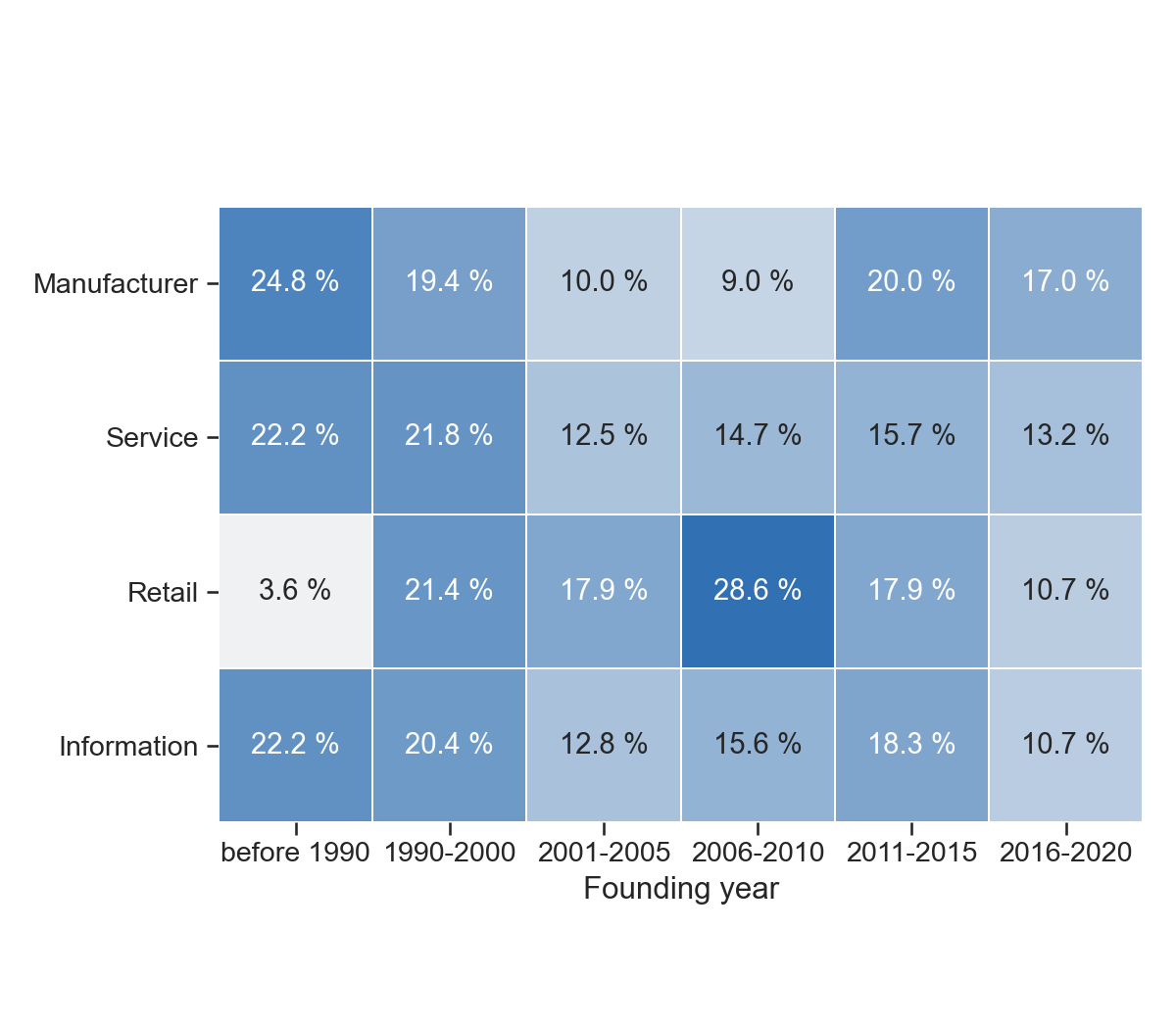}
    \caption{Age of 3D Printing Companies}
    \label{fig:3D_Companies_Age}
\end{figure}

Distinguishing engaged firms of different sizes, we see that small (10-49~employees) and micro firms (1-9~employees) account for the highest shares in all four classes. This was to be expected, given that the overwhelming majority of firms fall into this category. The highest share of large firms ($>$250~employees) is in manufacturing and the largest share of micro-sized firms is in retail (see Figure~\ref{fig:3D_Companies_Size}). 

\begin{figure}
    \centering
    \includegraphics[scale=0.43]{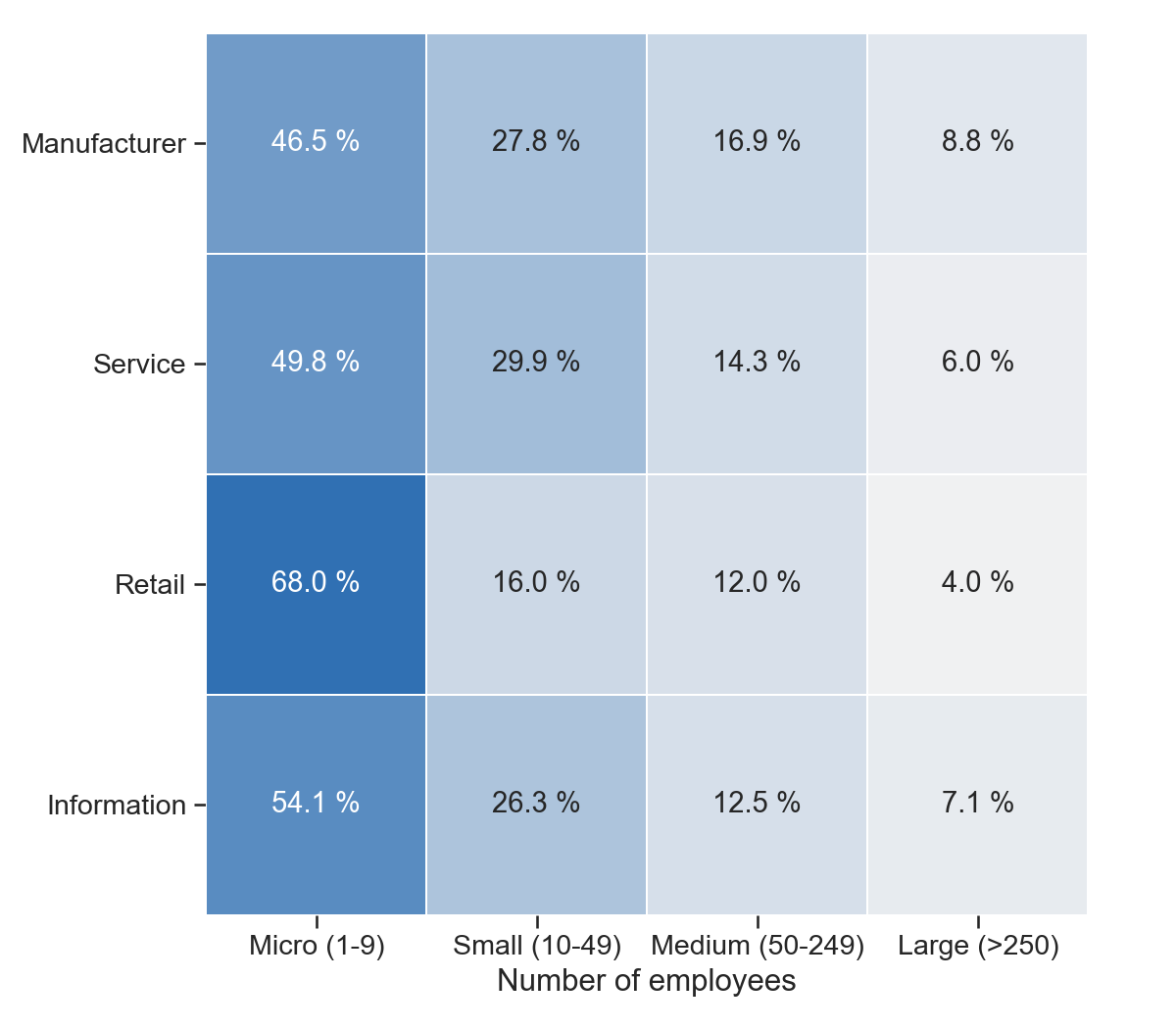}
    \caption{Size of 3D Printing Companies}
    \label{fig:3D_Companies_Size}
\end{figure}

Figure~\ref{fig:3D_companies_sector_distribution} provides a detailed insight into which industrial sectors 3D printing technology engagement is most prominent in.\footnote{See Table~\ref{tab:sector_classification} in the Appendix for details on the sector classification. We summarized sub-sectors into 31~main categories aggregating related activities. For instance, the category 'Electronics / optics' refers to NACE class C26 and covers manufacturing of electrical equipment including electric motors, machinery and equipment, appliances.} The share of 3D printing engaged firms is highest in material development such as synthetics (5.4\%) and metal (4.8\%). Relatively high shares can also be seen in sectors producing electronics and optical products (3.2\%) as well as in mechanical engineering (2.9\%) and chemicals (2.6\%). 
Some sectors are mainly active in 3D printing service provision such as the pharmaceutical industry (100\%), repair and installation services (89.5\%), and metal manufacturing (87.9\%). This finding is in line with the observation that 3D printing technology applications can be found in producing spare parts, appliances, medical devices, and drugs \cite{Pose-Rodriguez2020}.  
The largest relative proportion of manufacturers can be found in mining (25.0\%), textile/clothing (18.8\%), and transportation manufacturing (20.0\%). 
The sectors that show high shares of information providers related to 3D printing are the media \& publishing companies (56.1\%) as well as interest groups (45.7\%). This may reflect the value of information diffusion as a mean to reduce adoption barriers for end users.  
The prevalence of 3D printing engagement in the retail category is highest in textile/clothing (6.2\%) and unsurprisingly in wholesale (1.8\%) and retailing itself (1.7\%).
 
\subsection{Geographical diffusion}
Figure~\ref{fig:Germany_Map_Distribution_National} shows the location of the identified firms engaged in 3D printing technology using the listed address in ORBIS as the location identifier. We see an agglomeration of engaged companies in larger cities such as Berlin, Munich, and Hamburg. Yet, we also find relatively high absolute numbers across North-Rhine Westphalia, particularly in the area around Cologne and in the south-west of Germany (Baden-Württemberg). The high overall agglomeration near Cologne is likely based on the diverse sectors located in the region ranging from mining and consumer products to clothing and design. 
Moreover, the areas around Hanover and Dresden stand out as having relatively many engaged companies. We assume that this geographical pattern reflects the overall business activities in the vicinity of active locations. For example, the appearance of Bremen and Kiel may reflect the presence of ship manufacturing, whereas in the south there is a stronger focus on automotive manufacturing. \\  
Even more informative, however, is the depiction in Figure~\ref{fig:Germany_Map_Distribution_Regional} that shows the number of 3D active firms relative to the overall number of firms in the same region. In this visualization, it becomes more evident that the relative intensity is highest in the south and south-west. In addition, there are adoption hotspots in central Germany and in some areas in the north. 
The red squares indicate the top~10 locations in terms of relative 3D engagement intensity. \\   
The three most engaged regions are Jena (Thuringia), Tuttlingen (Baden-Württemberg), and Starnberg (Bavaria). Six out of the ten biggest clusters are located either in Baden-Württemberg or in Bavaria. Only two of the hotspots are not located in southern Germany (Olpe and Jena).
Taking a closer look at the area around Munich (see Figure~\ref{fig:munichHeatmap}), we find several micro hotspots within and around the city. Starnberg is one of top~10 hotspots and at the same time the location of the headquarter of EOS - one of the largest manufacturers of 3D printers and equipment. Located in the south and south-east of Munich are several other high-tech companies including Airbus and Infineon Technologies.\\
Moreover, within the city district of Munich is the headquarter of the car manufacturer BMW. Other big corporations such as Siemens AG (among the top~3 applicants for 3D printing patents) and MTU Aero Engines (among the top~10 patent applicants) are located in this area (cf. \cite{Pose-Rodriguez2020}, Table~4.1).\\
Jena, a comparatively much smaller city than Munich, is a location where the engagement might be driven by users from the optical industry, which has been traditionally very strong in this region.\\
Tuttlingen and Pforzheim as other high engagement locations, on the other hand, have been traditionally strong in manufacturing sectors such as suppliers for the automotive sectors, medical engineering, or the construction of parts and materials used in industrial production including jewellery. A look at the overall distribution of manufacturing firms (see Figure~\ref{fig:Germany_Map_Distribution_Manufacturing}) suggests that diffusion and adoption seem to be related to the prevalence of traditional manufacturing firms. This hypothesis is supported by the hotspot Frankenthal (Pfalz), located close to Ludwigshafen am Rhein, where the chemical corporation BASF, a holder of several 3D printing patents \cite{Pose-Rodriguez2020}, has its headquarters.

\begin{figure*}[h!]
    \centering
    \includegraphics[scale=0.35]{figures/type_sectors_tum_style_en_v3_DE.png}
    \caption{Sector Distribution of 3D Printing Companies}
    \label{fig:3D_companies_sector_distribution}
\end{figure*}

Another pattern that becomes visible in Figure~\ref{fig:Germany_Map_Distribution_Universities} is the presence of technical universities (TU) in and close to highly engaged locations. This is also visible in Figure~\ref{fig:munichHeatmap} where the most prominent downtown hotspot is the location of the Technical University of Munich (TUM). In addition, the Munich University of the Federal Armed Forces (\textit{Bundeswehruniversität}) is located in the south-east of the city where we also observe high-intensity zones. \\
This pattern is not unique to Munich, but can also be seen in Karlsruhe and Aachen, related to the Karlsruhe Institute of Technology (KIT) and the RWTH Aachen, respectively. Moreover, there are plenty of smaller universities of applied sciences, such as the Coburg University of Applied Science and Arts, in less densely populated areas that may also contribute to relatively high adoption intensities in their surroundings.\\ 
A more differentiated analysis can be made when comparing the four types \textit{Manufacturer}, \textit{Service}, \textit{Retailer}, and \textit{Information}. Regions with a higher density of traditional manufacturing firms show a stronger engagement in manufacturing (see Figure~\ref{fig:Germany_Map_labels_manufacturer}) and retail (see Figure~\ref{fig:Germany_Map_labels_retail}). Locations in close proximity to technical universities are stronger in 3D printing as a service (see Figure~\ref{fig:Germany_Map_labels_service}) and information provision (see Figure~\ref{fig:Germany_Map_labels_information}). The latter seems to take place in locations with a strong manufacturing base such as Lichtenfels, where General Electric owns a majority share in the company Concept Laser and has since expanded its activities and attracts suppliers and service providers to the region. In the identified top~10 locations, we see all types being active which suggests that clusters emerge when there is both manufacturing and applications as well as information provision. Yet, there are areas with relatively high engagement intensity that are rather active in retail. We find agglomerations north and south of Berlin as well as in the north of Germany suggesting that inexpensive storage space and accessibility to export markets may determine the location choice.  

\begin{figure*}
    \centering
    \includegraphics[scale=0.7]{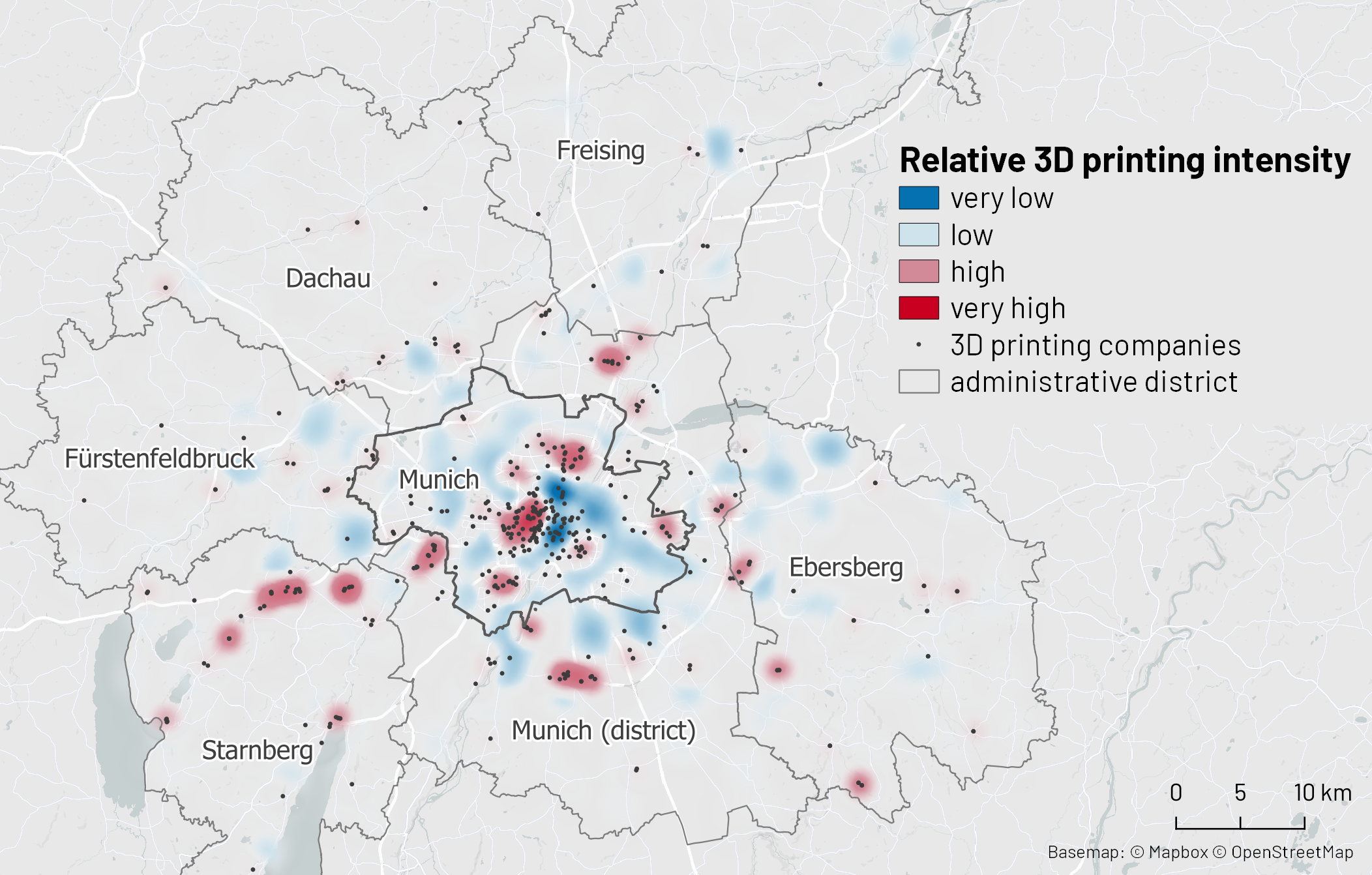}
    \caption{3D Printing Intensity in Munich}
    \label{fig:munichHeatmap}
\end{figure*}

\subsection{Validation}
To check the validity of the webAI type identification, we perform a consistency test that checks the overall innovativeness of the firms by category. If our identification of new technology adopters and diffusers worked, we would expect that those firms are also more innovative. We make use of the InnoProb, a measure for firms overall innovativeness as developed in \cite{Kinne2021}, which assigns an innovation probability to firms based on a model trained with Community Innovation Survey data. As can be seen in Figure~\ref{fig:3D_Companies_Innovativeness}, the 3D printing technology engagement correlates with the degree of innovativeness of the respective types. We found that 57.7\% of companies engaged in 3D printing were innovative (InnoProb $\geq 0.4$). For manufacturers, as many as 74.9\% were innovative companies. This suggests that we were capturing firms that have recently introduced a service or product innovation. This is true for all four types, but particularly for information providers and manufacturers.     
Our results regarding the use of 3D printing by sector are in line with the findings presented by Pose-Rodriguez et al. \cite{Pose-Rodriguez2020}. This report is based on patent application information (as recorded in the PATSTAT database) and documents trends in patenting by sector, which also confirms the prevalence of 3D printing in transport manufacturing, industrial tooling, health, optics applications, and consumer goods. Our mapping also overlaps geographically with some of the findings in \cite{Pose-Rodriguez2020} (see for instance Figure~38 in their paper), but also identified additional regions not found in patent data. There is also a significant overlap in the identified hotspots, as the report also found Munich, Munich district, Berlin, Hamburg and Starnberg. Yet, patent screening alone may underestimate the role of locations like Jena or regions that are less manufacturing-intensive. 

\begin{figure}
    \centering
    \includegraphics[scale=0.48]{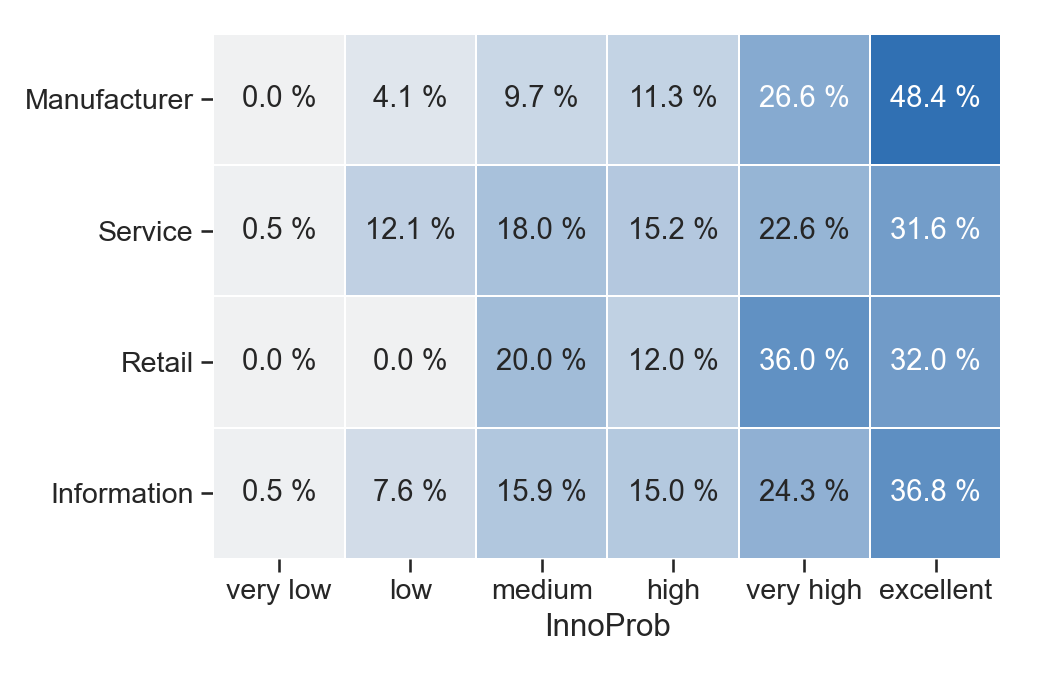}
    \caption{Innovation Performance of 3D Printing Companies}
    \label{fig:3D_Companies_Innovativeness}
\end{figure}

\section{Discussion and Conclusion}
This study illustrates a website text-based approach to analyze new technology adoption. WebAI utilizes machine learning to identify companies that engage in specific technologies which are not easily reflected in traditional indicators such as patents. Moreover, it allows to differentiate between the type of engagement of companies capturing also activities that are related to the diffusion of new technology rather than its invention. Such novel indicators may be informative for policy makers as well as other stakeholders by providing additional information about individual technologies.\\   
The mapping for the case of 3D printing revealed interesting geographical patterns of technology use and diffusion. Linking adoption intensity to the characteristics of the companies shows that young and small firms play an important role. Furthermore, we show that regions with a strong general manufacturing industry have higher levels of activities. Interestingly, the great majority of 3D printing-engaged companies is classified as the type \textit{Service}. Locations with technical universities show higher adoption intensities suggesting that academic research organizations and basic research may play an important role for pushing new technologies and the provision of skilled employees trained in relevant fields. It should be noted, however, that the presented analysis is merely descriptive. Subsequent analyses may use econometric models to uncover statistically significant relationships. Moreover, the analysis of causal relationships between location factors, such as universities, and 3D-printing technology adoption would be interesting and could augment recent research on that topic \cite{Hausman2021}. Comparing the results to analyses based on patent data could furthermore inform innovation research and indicator development. Finally, a deeper investigation of the link between webAI-based measures and other innovation indicators at the company-level would be desirable for a better understanding of the potentially different dimensions captured by both approaches for mapping technology diffusion \cite{Kinne2020}. \\

We encourage more research on the application of webAI to other technologies as well as the validation of our results by other researchers. We made some efforts to validate our results internally as well as externally, but we acknowledge that the extent to which our labels can be generalized may be limited. The accuracy was significantly influenced by how consistently the training data set could be labeled in the first step, i.e. by how evident the allocation of each data point to the respective cluster was for each labelling person. Moreover, while relying on company websites has the advantage of almost universal coverage, there may be very small or young companies that do not have a website (yet) and are therefore not included in our analysis \cite{Kinne2021}. Finally, it should be noted that the websites were analyzed at a fixed point in time and that statements about adoption dynamics will require repeating the analyses in the future.

\begin{figure*}
    \centering
    \begin{subfigure}[b]{0.4\textwidth}
    \centering
    \includegraphics[width=\textwidth]{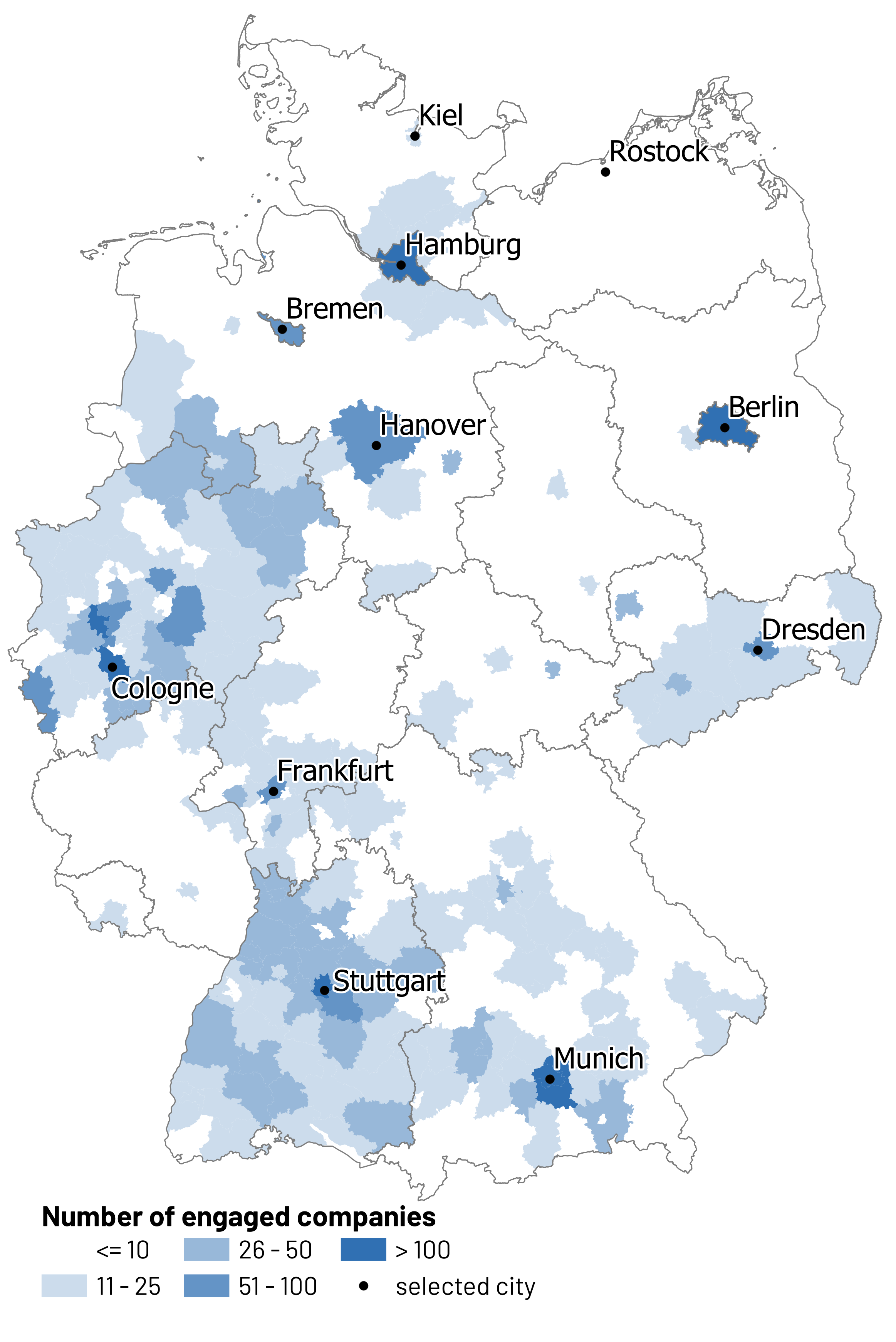}        \caption[]%
            {{\small Absolute Intensity}}    
    \label{fig:Germany_Map_Distribution_National}
        \end{subfigure}
        \hfill
        \begin{subfigure}[b]{0.4\textwidth}
    \centering 
    \includegraphics[width=\textwidth]{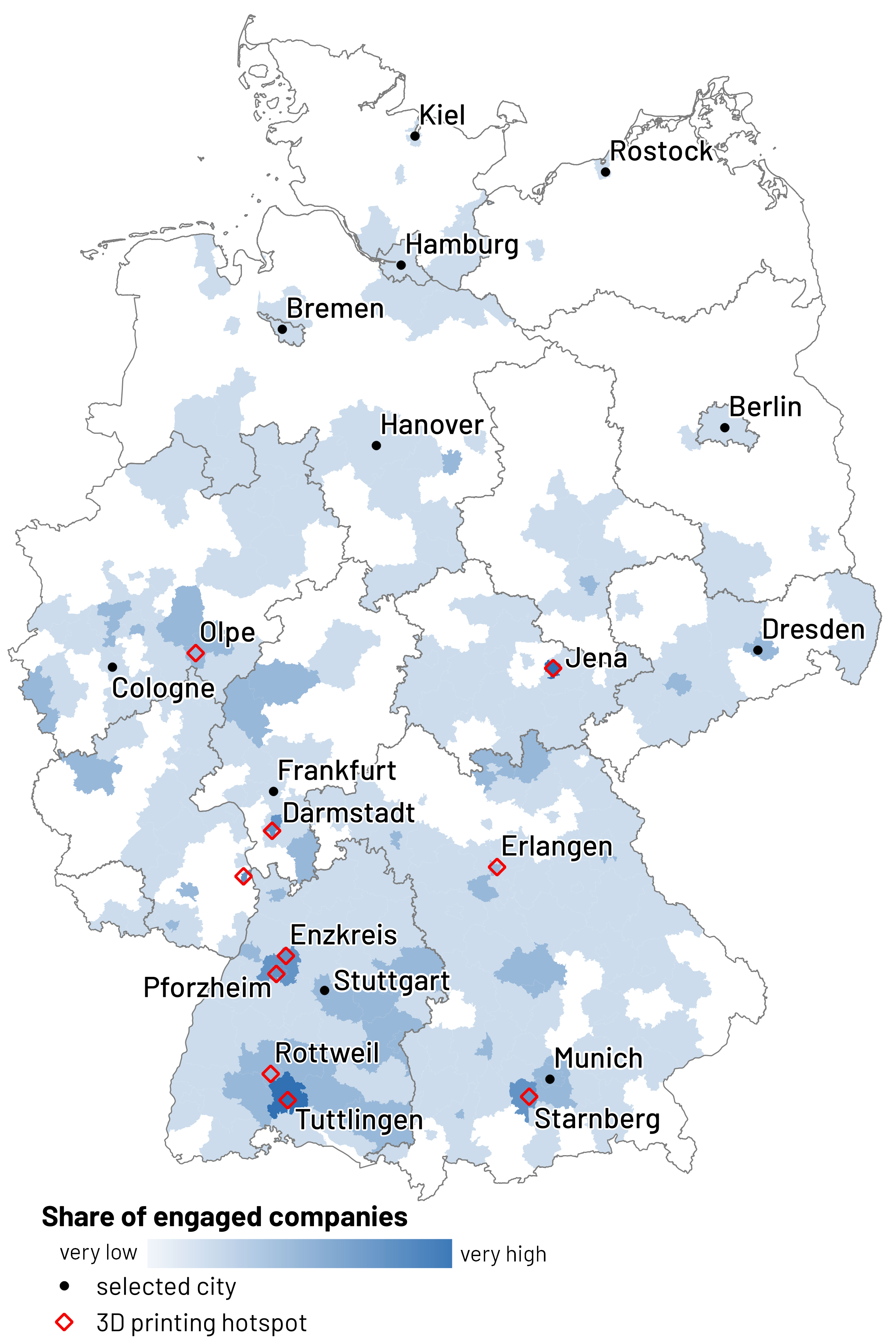}
        \caption[]%
            {{\small Relative Intensity}}    
    \label{fig:Germany_Map_Distribution_Regional}
    \end{subfigure}
    \vskip\baselineskip
    \begin{subfigure}[b]{0.4\textwidth} 
        \centering 
        \includegraphics[width=\textwidth]{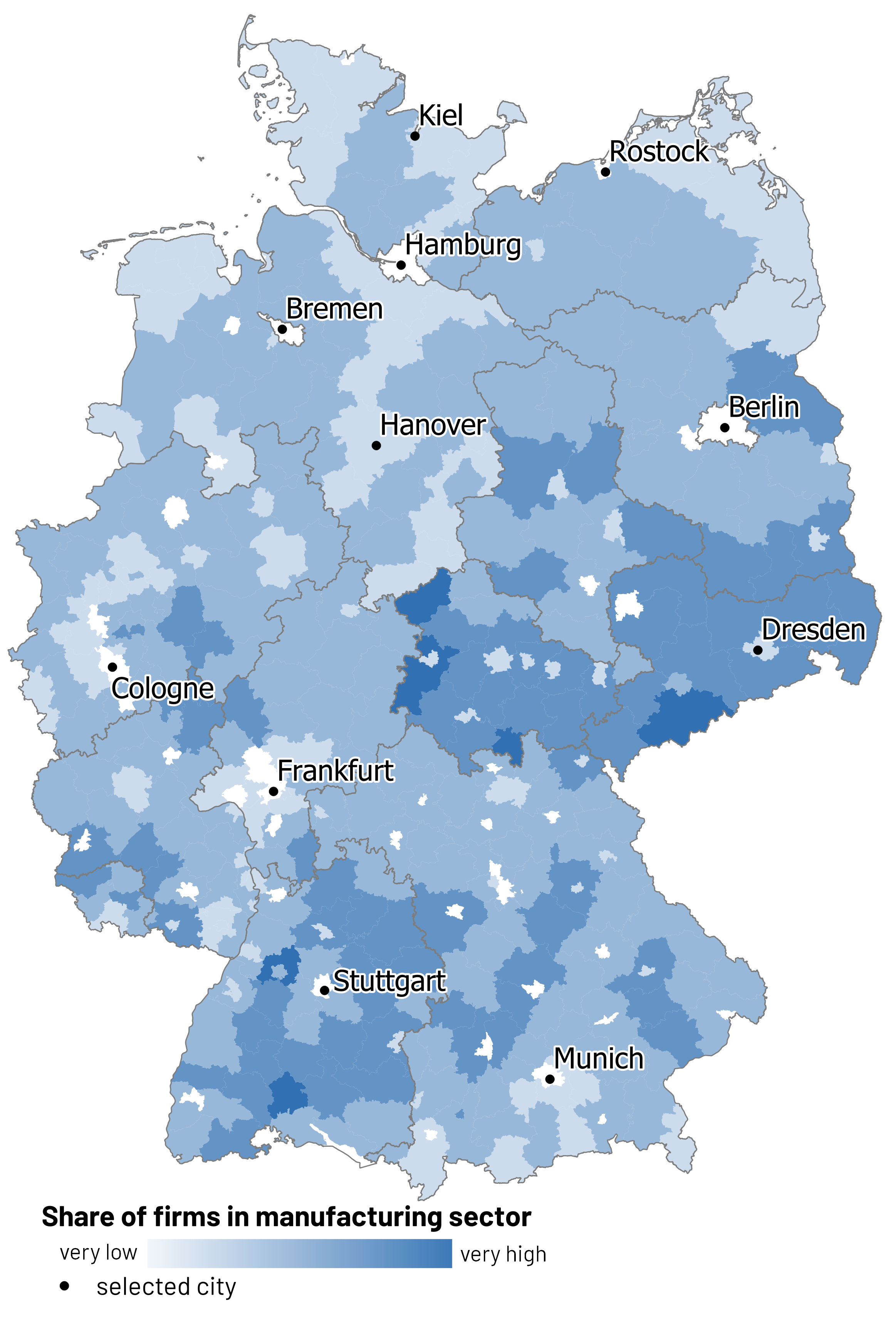}
        \caption[]%
        {{\small Total Manufacturing Industry}}    
        \label{fig:Germany_Map_Distribution_Manufacturing}
    \end{subfigure}
    \hfill
    \begin{subfigure}[b]{0.4\textwidth} 
        \centering 
        \includegraphics[width=\textwidth]{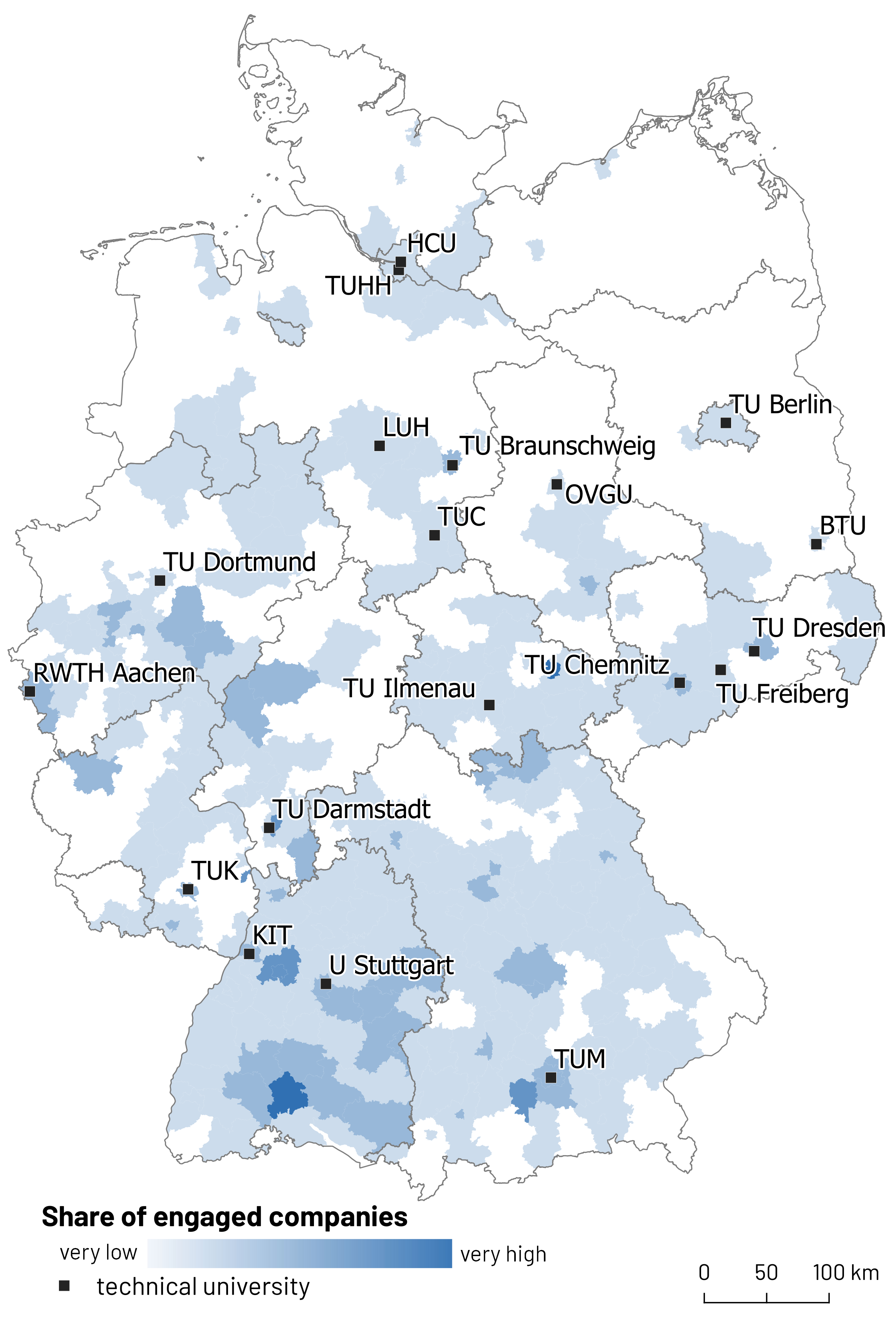}
        \caption[]%
        {{\small Technical Universities}}    
        \label{fig:Germany_Map_Distribution_Universities}
    \end{subfigure}
    \caption[Heat maps of Germany] {\small 3D Printing Intensity and Facilitators } 
    \label{fig:Germany_Map_Distribution_Total}
\end{figure*}

\begin{figure*}
    \centering
    \begin{subfigure}[b]{0.4\textwidth}
    \centering
    \includegraphics[width=\textwidth]{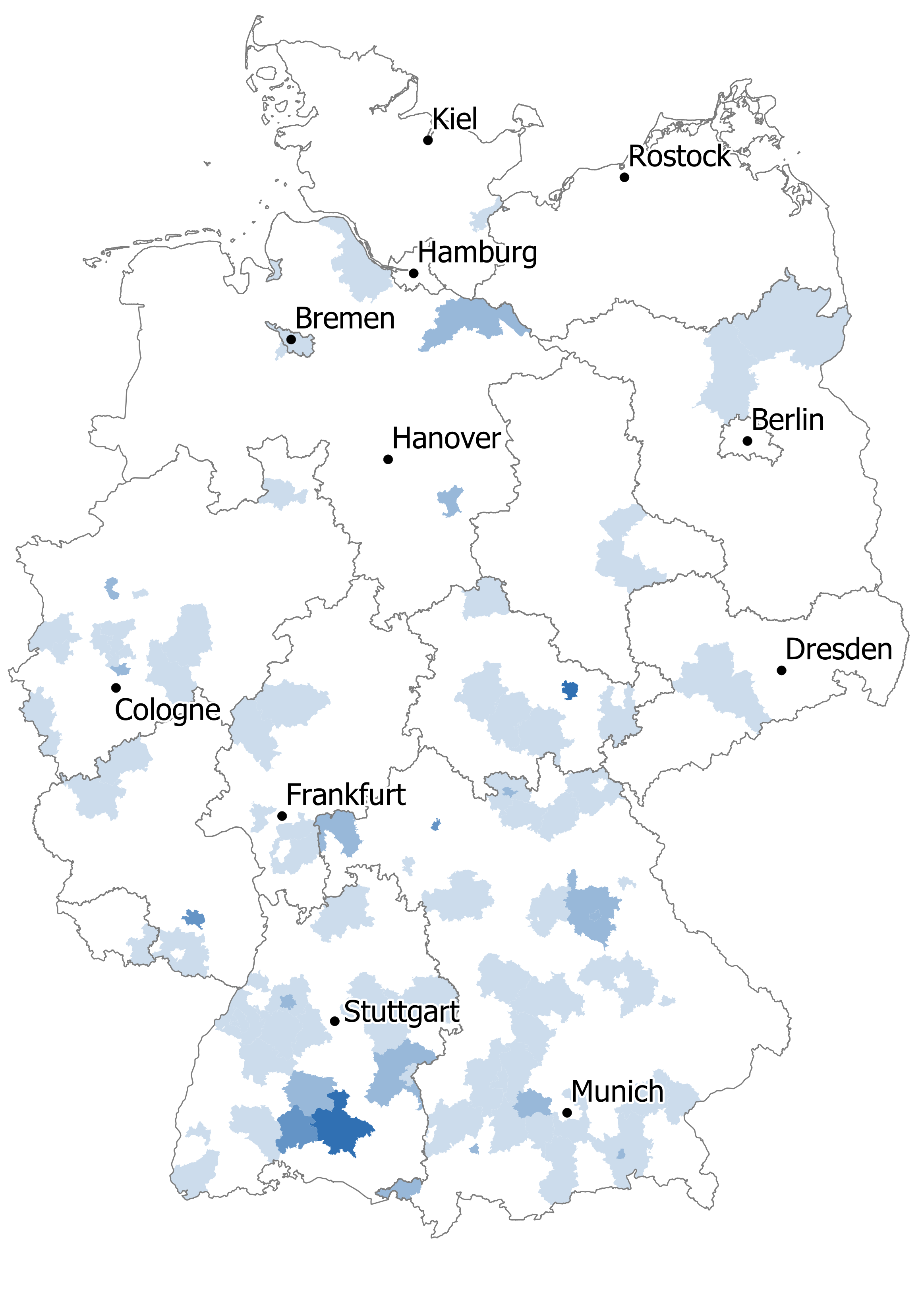}        \caption[Network2]%
            {{\small Manufacturer}}    
    \label{fig:Germany_Map_labels_manufacturer}
        \end{subfigure}
        \hfill
        \begin{subfigure}[b]{0.4\textwidth}
    \centering 
    \includegraphics[width=\textwidth]{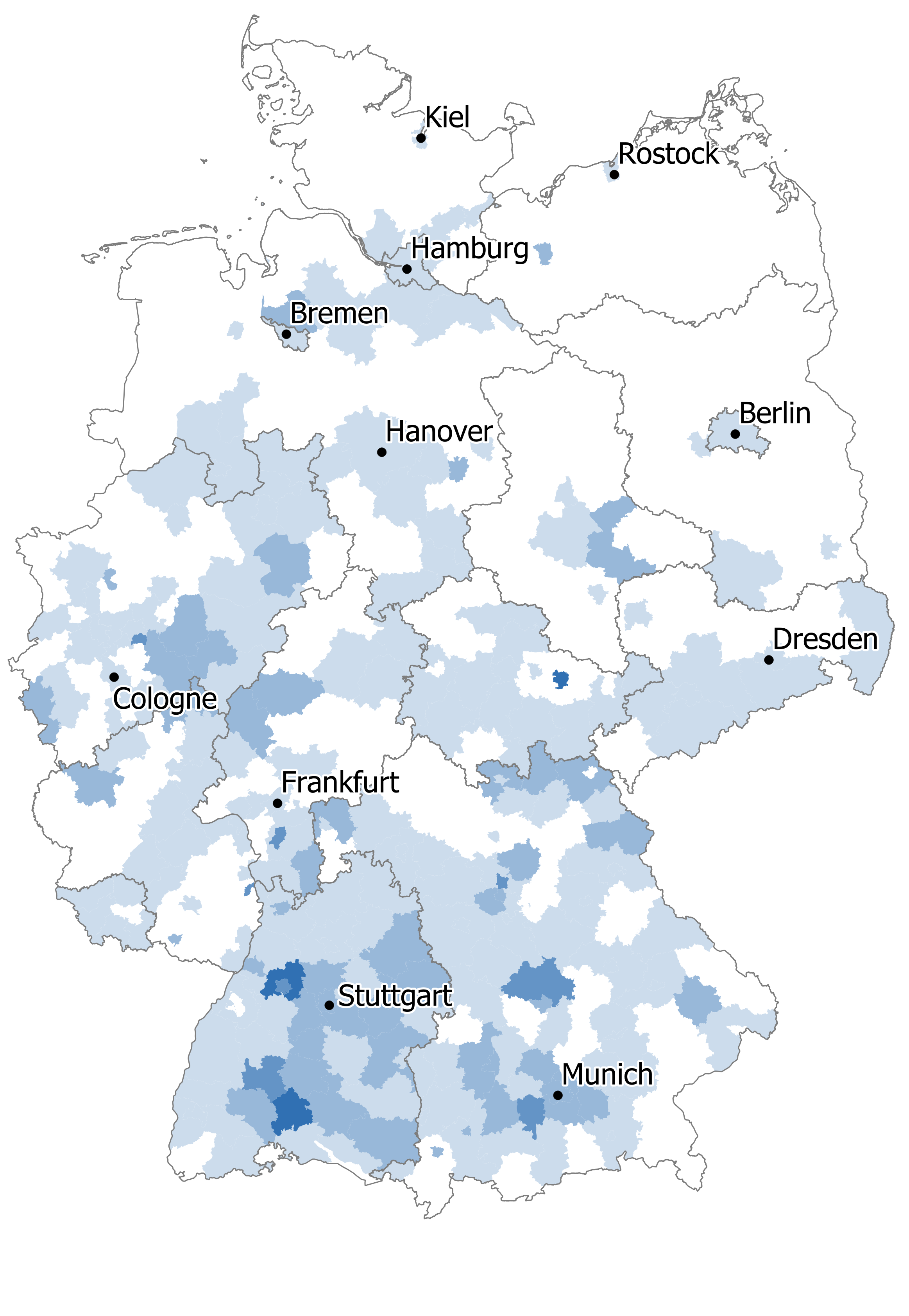}
        \caption[]%
            {{\small Service Provider}}    
    \label{fig:Germany_Map_labels_retail}
    \end{subfigure}
    \vskip\baselineskip
    \begin{subfigure}[b]{0.4\textwidth}   
        \centering 
        \includegraphics[width=\textwidth]{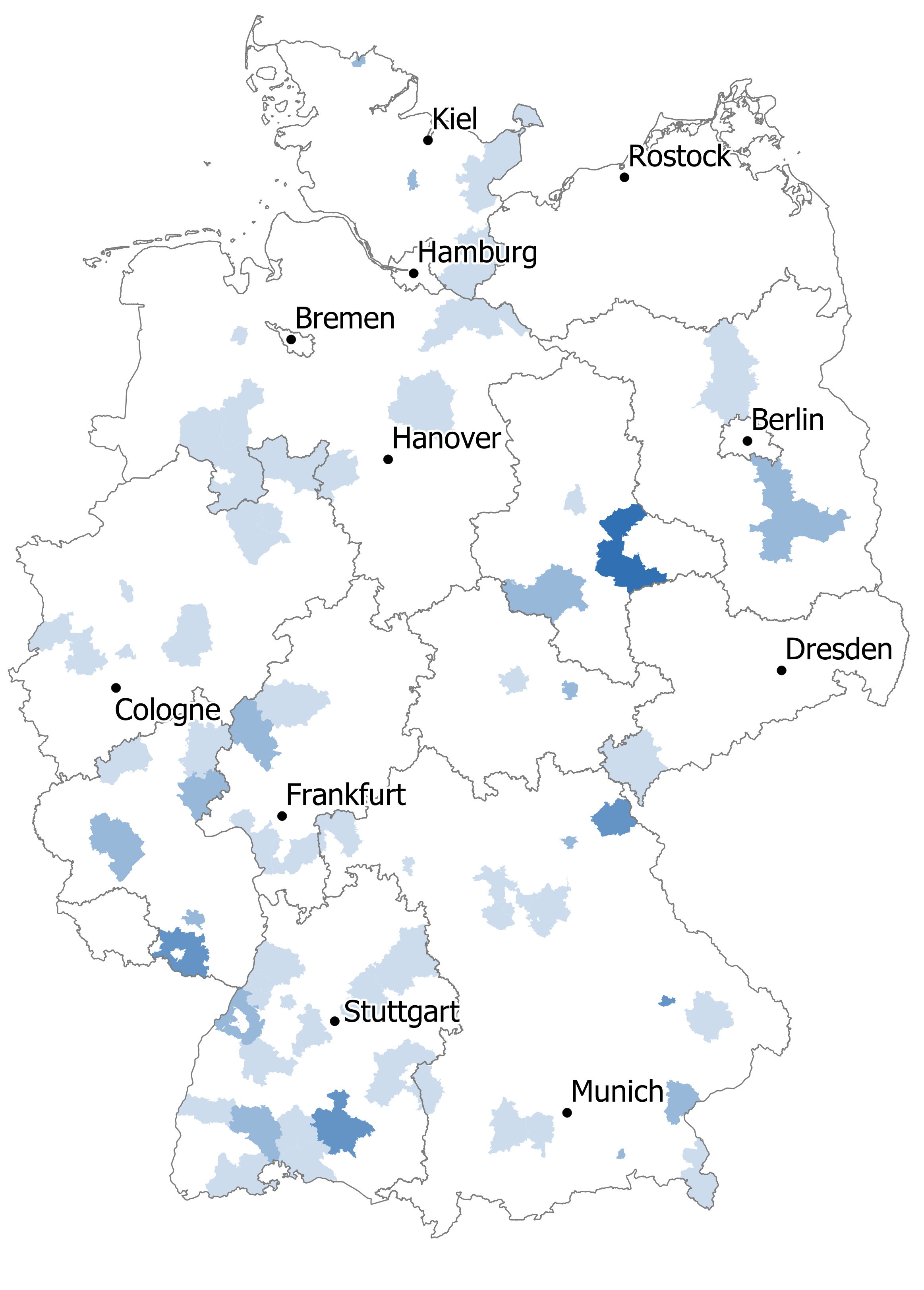}
        \caption[]%
        {{\small Retail}}    
        \label{fig:Germany_Map_labels_service}
    \end{subfigure}
    \hfill
    \begin{subfigure}[b]{0.4\textwidth} 
        \centering 
        \includegraphics[width=\textwidth]{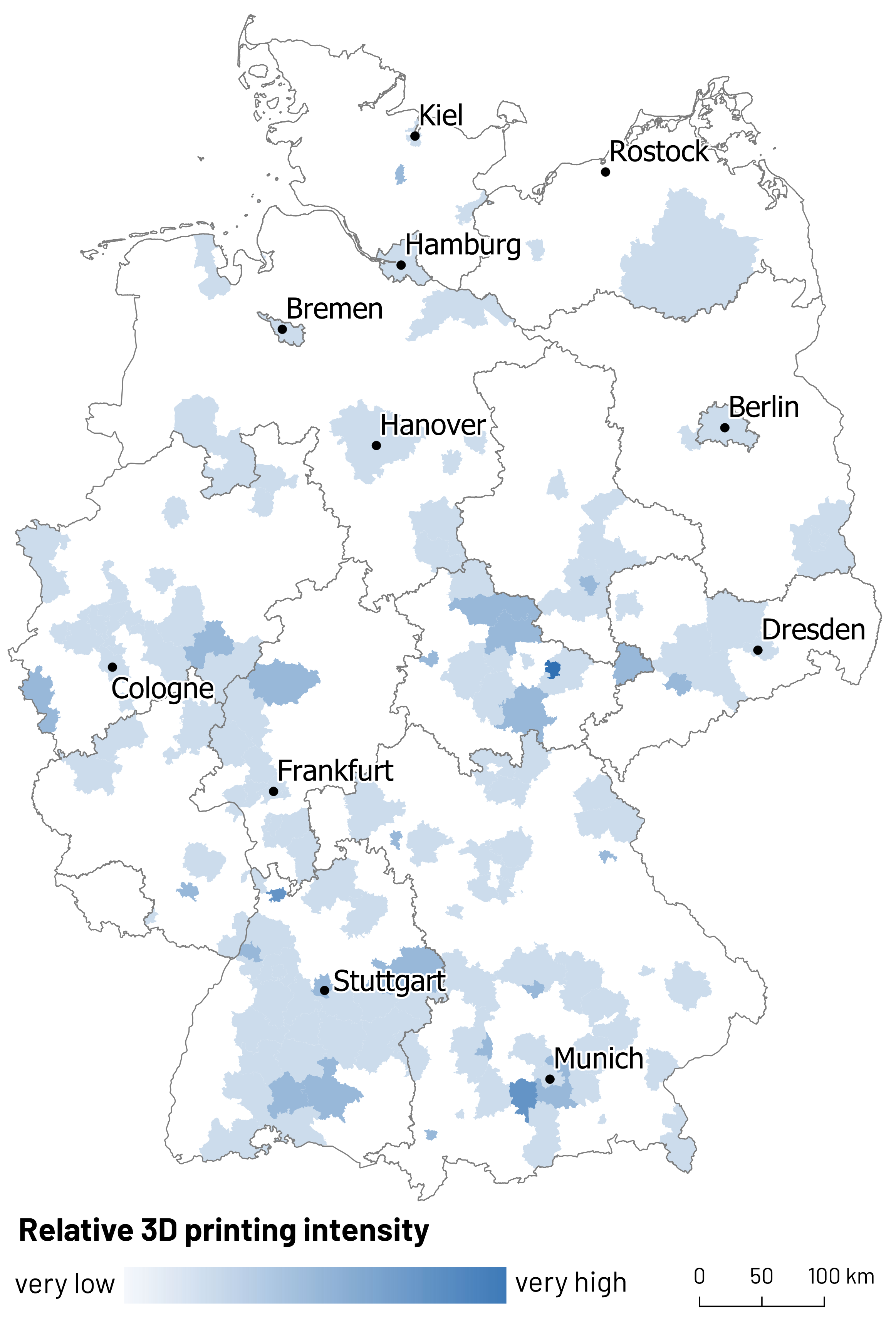}
        \caption[]%
        {{\small Information}}    
        \label{fig:Germany_Map_labels_information}
    \end{subfigure}
    \caption[Relative 3D printing intensity differentiated by label categories ]
    {\small Relative 3D Printing Intensity Differentiated by Categories} 
    \label{fig:Germany_Map_labels_total}
\end{figure*}

\clearpage
\bibliographystyle{plain}
\bibliography{references}

\begin{thebibliography}{10}

\bibitem{AghionAntoninBunel+2021}
Philippe Aghion, Céline Antonin, and Simon Bunel.
\newblock {\em The Power of Creative Destruction: Economic Upheaval and the
  Wealth of Nations}.
\newblock Harvard University Press, 2021.

\bibitem{Aghion2009}
Philippe Aghion, Richard Blundell, Rachel Griffith, Peter Howitt, and Susanne
  Prantl.
\newblock {The Effects of Entry on Incumbent Innovation and Productivity}.
\newblock {\em The Review of Economics and Statistics}, 91(1):20--32, 2009.

\bibitem{Audretsch1996}
David~B. Audretsch and Maryann~P. Feldman.
\newblock {R\&D} spillovers and the geography of innovation and production.
\newblock {\em The American Economic Review}, 86(3):630--640, 1996.

\bibitem{ben2017decentralization}
Avner Ben-Ner and Enno Siemsen.
\newblock Decentralization and localization of production: The organizational
  and economic consequences of additive manufacturing (3d printing).
\newblock {\em California Management Review}, 59(2):5--23, 2017.

\bibitem{Cohen2002}
Wesley~M. Cohen, Richard Nelson, and John~P. Walsh.
\newblock Links and impacts: The influence of public research on industrial
  {R\&D}.
\newblock {\em Management Science}, 48(1):1--23, 2002.

\bibitem{Crepon1998}
Bruno Crepon, Emmanuel Duguet, and Jacques Mairesse.
\newblock {Research, Innovation And Productivity: An Econometric Analysis At
  The Firm Level}.
\newblock {\em Economics of Innovation and New Technology}, 7(2):115--158,
  1998.

\bibitem{Eyers2018}
Daniel~R. Eyers, Andrew~T. Potter, Jonathan Gosling, and Mohamed~M. Naim.
\newblock {The flexibility of industrial additive manufacturing systems}.
\newblock {\em International Journal of Operations and Production Management},
  38(12):2313--2343, dec 2018.

\bibitem{gebhardt2016additive}
Andreas Gebhardt and Jan-Steffen H{\"o}tter.
\newblock {\em Additive manufacturing: 3D Printing for Prototyping and
  Manufacturing}.
\newblock Carl Hanser Verlag GmbH \& Co. KG, Munich, 2016.

\bibitem{ghobadian2020examining}
Abby Ghobadian, Irene Talavera, Arijit Bhattacharya, Vikas Kumar, Jose~Arturo
  Garza-Reyes, and Nicholas O'regan.
\newblock Examining legitimatisation of additive manufacturing in the interplay
  between innovation, lean manufacturing and sustainability.
\newblock {\em International Journal of Production Economics}, 219:457--468,
  2020.

\bibitem{Hall2006}
Bronwyn~H. Hall.
\newblock {\em Innovation and Diffusion}, chapter~17, pages 312--326.
\newblock Oxford University Press, Oxford, UK, 2006.

\bibitem{Hall2020}
Bronwyn~H. Hall.
\newblock Patents, innovation, and development.
\newblock {\em Max Planck Institute for Innovation \& Competition Research
  Paper}, No. 20-07, 2020.

\bibitem{Hausman2021}
Naomi Hausman.
\newblock {University Innovation and Local Economic Growth}.
\newblock {\em The Review of Economics and Statistics}, pages 1--46, 03 2021.

\bibitem{ISO/ASTM2018}
ISO/ASTM.
\newblock {\em Additive manufacturing - General principles - Terminology
  (ISO/ASTM DIS 52900:2018)}.
\newblock Beuth Verlag, Berlin, 2018.

\bibitem{jin2019auto}
Haifeng Jin, Qingquan Song, and Xia Hu.
\newblock Auto-keras: An efficient neural architecture search system.
\newblock In {\em Proceedings of the 25th ACM SIGKDD International Conference
  on Knowledge Discovery \& Data Mining}, pages 1946--1956. ACM, 2019.

\bibitem{Kinne2020}
Jan Kinne and Janna Axenbeck.
\newblock {Web mining for innovation ecosystem mapping: a framework and a
  large-scale pilot study}.
\newblock {\em Scientometrics}, 125, 2020.

\bibitem{Kinne2021}
Jan Kinne and David Lenz.
\newblock Predicting innovative firms using web mining and deep learning.
\newblock {\em PLoS ONE}, 16(4):e0249071, 2021.

\bibitem{malte2019evolution}
Aditya Malte and Pratik Ratadiya.
\newblock Evolution of transfer learning in natural language processing.
\newblock {\em arXiv preprint arXiv:1910.07370}, 2019.

\bibitem{MOSER2013}
Petra Moster.
\newblock Patents and innovation: Evidence from economic history.
\newblock {\em Journal of Economic Perspectives}, 27:23--44, 2013.

\bibitem{Pose-Rodriguez2020}
Javier Pose-Rodriguez, Judy Ceulemans, Yann Ménière, Aliki Nichogiannopoulou,
  and Ilja Rudyk.
\newblock Patents and additive manufacturing: Trends in 3d printing
  technologies.
\newblock 07 2020.

\bibitem{reimers2019sentencetransformers}
Nils Reimers and Iryna Gurevych.
\newblock Sentence-{BERT}: Sentence embeddings using {S}iamese {BERT}-networks.
\newblock In {\em Proceedings of the 2019 Conference on Empirical Methods in
  Natural Language Processing and the 9th International Joint Conference on
  Natural Language Processing (EMNLP-IJCNLP)}, pages 3982--3992, Hong Kong,
  China, November 2019. Association for Computational Linguistics.

\bibitem{Salazar2004}
Mónica Salazar and Adam Holbrook.
\newblock {A debate on innovation surveys}.
\newblock {\em Science and Public Policy}, 31(4):254--266, 08 2004.

\bibitem{schwierzy2021digitalisation}
Julian Schwierzy.
\newblock Digitalisation of production: Industrial additive manufacturing and
  its implications for competition and social welfare.
\newblock {\em Munich Papers in Political Economy}, 16, 2021.

\bibitem{su2018history}
Amanda Su and Subhi~J Al'Aref.
\newblock History of 3d printing.
\newblock In {\em 3D Printing Applications in Cardiovascular Medicine}, pages
  1--10. Elsevier, 2018.

\bibitem{weller2015economic}
Christian Weller, Robin Kleer, and Frank~T Piller.
\newblock Economic implications of 3d printing: Market structure models in
  light of additive manufacturing revisited.
\newblock {\em International Journal of Production Economics}, 164(C):43--56,
  2015.

\end{thebibliography}
\onecolumn
\clearpage

\setcounter{figure}{0}
\setcounter{table}{0}
\renewcommand{\thetable}{A\arabic{table}}
\renewcommand{\thefigure}{A\arabic{figure}}
\section*{Appendix}

\footnotesize
\begin{longtable}{L{6cm} L{5cm} L{5.5cm}}
   \caption{3D Printing Keywords}\\

    \textbf{English} & \textbf{German} & \textbf{Source}\\
    \hline
Additive Manufacturing                         & Additive Fertigung              & Norm: VDI 3405                    \\
Rapid Manufacturing                             & -                               & Norm: VDI 3405                     \\
Rapid Prototyping\textsuperscript{\textdied{}}                              & -                               & Norm: VDI 3405                     \\
Rapid Tooling                                 & -                               & Norm: VDI 3405                     \\
Stereolithography                              & Stereolithografie               & Norm: VDI 3405                     \\
Laser Sintering                                & Laser-Sintern                   & Norm: VDI 3405                     \\
Selective Laser Sintering                      & Selektives Laser-Sintern        & Norm: VDI 3405                     \\
Laser Beam Melting                             & Laser-Strahlschmelzen           & Norm: VDI 3405                     \\
Laser Forming                                  & Laser Forming                   & Norm: VDI 3405                     \\
Selective Laser Melting                        & Selective Laser Melting         & Norm: VDI 3405                     \\
LaserCUSING                                    & LaserCUSING                     & Norm: VDI 3405                     \\
Direct Metal Laser Sintering                   & Direktes Metall-Laser-Sintern   & Norm: VDI 3405                     \\
Electron Beam Melting                          & Elektronen-Strahlschmelzen      & Norm: VDI 3405                     \\
Fused Layer Manufacturing                      & Fused Layer Manufacturing       & Norm: VDI 3405                     \\
Fused Layer Modelling                          & Fused Layer Modelling           & Norm: VDI 3405                     \\
Fused Deposition Modelling                     & Fused Deposition Modelling      & Norm: VDI 3405                     \\
Filament Deposition                            & Strangablegeverfahren           & Norm: VDI 3405                     \\
Multi-Jet Modelling                            & Multi-Jet Modelling             & Norm: VDI 3405                     \\
Poly-Jet Modelling                             & Poly-Jet Modelling              & Norm: VDI 3405                     \\
3D-Printing                                    & 3D-Druck                        & Norm: VDI 3405                     \\
Layer Laminated Manufacturing                  & Schicht-Laminat-Verfahren       & Norm: VDI 3405                     \\
Laminated Object Manufacturing                 & Laminated Object Manufacturing  & Norm: VDI 3405                     \\
Digital Light Processing                       & Digital Light Processing        & Norm: VDI 3405                     \\
Thermotransfer Sintering                       & Thermotransfer-Sintern          & Norm: VDI 3405                     \\
Binder Jetting                                 & Freistrahl-Bindemittelauftrag   & Norm: ASTM 52900                   \\
Directed Energy Deposition               & Materialauftrag mit gerichteter Energieeinbringung & Norm: ASTM 52900      \\
Material Extrusion                             & Materialextrusion               & Norm: ASTM 52900                   \\
Material Jetting                               & Freistrahl-Materialauftrag      & Norm: ASTM 52900                   \\
Powder Bed Fusion                              & Pulverbettbasiertes Schmelzen   & Norm: ASTM 52900                   \\
Sheet Lamination                               & Schichtlaminierung              & Norm: ASTM 52900                   \\
Vat Photopolymerization                        & Badbasierte Photopolymerisation & Norm: ASTM 52900                   \\
Metal Selective Laser Sintering                & -                               & Consulting: AMPower                \\
Laser Beam Powder Bed Fusion                   & -                               & Consulting: AMPower                \\
Electron Beam Powder Bed Fusion                & -                               & Consulting: AMPower                \\
Powder Feed Laser Energy Deposition            & -                               & Consulting: AMPower                \\
Coldspray                                   & -                               & Consulting: AMPower                \\
Wire Arc \textsuperscript{\textdied{}}                                      & -                               & Consulting: AMPower                \\
Plasma Arc Energy Deposition                   & -                               & Consulting: AMPower                \\
Wire Feed Laser                                & -                               & Consulting: AMPower                \\
Electron Beam Energy Deposition                & -                               & Consulting: AMPower                \\
Resistance Welding \textsuperscript{\textdied{}}                           & -                               & Consulting: AMPower                \\
Liquid Metal Printing                          & -                               & Consulting: AMPower                \\
Ultrasonic Welding \textsuperscript{\textdied{}}                         & -                               & Consulting: AMPower                \\
Friction Deposition                            & -                               & Consulting: AMPower                \\
Nanoparticle Jetting                           & -                               & Consulting: AMPower                \\
Metal Filament Fused Deposition Modeling       & -                               & Consulting: AMPower                \\
Metal Pellet Fused Deposition Modeling         & -                               & Consulting: AMPower                \\
Metal Lithography                              & -                               & Consulting: AMPower                \\
Powder Metallurgy Jetting                      & -                               & Consulting: AMPower                \\
Mold Slurry Deposition                         & -                               & Consulting: AMPower                \\
Electrographic Sheet Lamination                & -                               & Consulting: AMPower                \\
Thermal Powder Bed Fusion                      & -                               & Consulting: AMPower                \\
Laser Powder Bed Fusion                        & -                               & Consulting: AMPower                \\
Pellet Based Material Extrusion                & -                               & Consulting: AMPower                \\
Continuous Fiber Thermoplastic   Deposition    & -                               & Consulting: AMPower                \\
Continuous Fiber Material Extrusion            & -                               & Consulting: AMPower                \\
Filament Based Material Extrusion              & -                               & Consulting: AMPower                \\
Continuous Fiber Sheet Lamination              & -                               & Consulting: AMPower                \\
Area-Wise Vat Polymerization                   & -                               & Consulting: AMPower                \\
Fiber Alignment Area-Wise Vat   Polymerization & -                               & Consulting: AMPower                \\
Thermoset Deposition                           & -                               & Consulting: AMPower                \\
Continuous Fiber Thermoset Deposition          & -                               & Consulting: AMPower                \\
Elastomer Deposition                           & -                               & Consulting: AMPower                \\
Vat Vulcanization                              & -                               & Consulting: AMPower                \\
Laser Chemical Vapor Deposition                & -                               & ISBN: 978-1-56990-582-1            \\
Continuous Liquid Interface Production         & -                               & ISBN: 978-1-56990-582-1            \\
Laser Metal Fusion                             & Selektives Laserschmelzen       & ISBN: 978-1-56990-582-1            \\
Selective Mask Sintering                       & Selektive Maskensintern         & ISBN: 978-1-56990-582-1            \\
Laser Engineered Net Shaping                   & Laserpulverformung              & ISBN: 978-1-56990-582-1            \\
ARBURG Kunststoff-Freiformen                   & -                               & ISBN: 978-1-56990-582-1            \\
Maskless Masoscale Material Deposition         & -                               & ISBN: 978-1-56990-582-1            \\
Foam Reaction Prototyping                      & -                               & ISBN: 978-1-56990-582-1            \\
Tool-Less Fabrication                       & Werkzeuglose Fertigung \textsuperscript{\textdied{}}       & DOI: 10.4337/9781781003930.00017   \\
Generative Manufacturing                       & Generative Fertigung            & DOI: 10.4103/jorr.jorr\_9\_17      \\
Digital Composite   Manufacturing              & -                               & DOI: 10.3233/AOP-120022            \\
-                                              & Laserstrahlschmelzen            & DOI: 10.25534/tuprints-00014474    \\
Laser Consolidation                            & Laserkonsolidierung             & DOI: 10.1533/9781845699819.6.492   \\
Ultrasonic Consolidation                 & Ultraschallkonsolidierung                          & DOI: 10.1177/0892705714565705           \\
Freeform Fabrication                           & Freiformherstellung             & DOI: 10.1109/JPROC.2016.2625103    \\
Layer Manufacturing                            & Schichtbauverfahren             & DOI: 10.1109/JPROC.2016.2625102    \\
Additive Layer Manufacturing             & Additive Schichtherstellung                        & DOI: 10.1109/JPROC.2016.2625101         \\
Additive Techniques \textsuperscript{\textdied{}}                           & Additive Techniken             & DOI: 10.1109/JPROC.2016.2625100    \\
Additive Process \textsuperscript{\textdied{}}                             & Additive Prozesse               & DOI: 10.1109/JPROC.2016.2625099    \\
Additive Fabrication                           & -                               & DOI: 10.1109/JPROC.2016.2625098    \\
Digital Light Synthesis                        & -                               & DOI: 10.1101/2020.05.25.20112201   \\
Continuous Digital Light   Manufacturing       & -                               & DOI: 10.1080/17452759.2012.673152  \\
Two-Photon Polymerization                      & -                               & DOI: 10.1063/1.4916886             \\
Hot Lithography                                & -                               & DOI: 10.1039/D0PY01746A            \\
Direct Shell Production Casting                & -                               & DOI: 10.1036/007142928X            \\
Thermal Polymerization                         & -                               & DOI: 10.1021/acsapm.8b00244        \\
Directed Light Fabrication                     & -                               & DOI: 10.1016/S0924-0136(97)00321-X \\
Light Initiated Fabrication   Technology & -                                                  & DOI: 10.1016/j.synthmet.2017.11.003     \\
Ultrasonic Additive Manufacturing        & Ultraschalladditivherstellung                      & DOI: 10.1016/j.scriptamat.2019.10.004   \\
E-Manufacturing \textsuperscript{\textdied{}}                              & -                               & DOI: 10.1016/j.pnsc.2008.03.027    \\
Programmable   Photopolymerization             & -                               & DOI: 10.1016/j.matdes.2020.109381  \\
Bioprinting                                    & -                               & DOI: 10.1016/j.ajps.2019.11.003    \\
Fused Filament Fabrication                     & Schmelzschichtung               & DOI: 10.1016/j.addma.2019.100861   \\
Masked Stereolithography                       & -                               & DOI: 10.1016/j.addma.2014.08.003   \\
Laser Metal Deposition                   & Laserauftragsschmelzen                             & DOI: 10.1016/B978-0-12-818411-0.00012-4 \\
Lithography-Based Ceramic Manufacturing      & -                               & DOI: 10.1007/s11665-017-2843-z     \\
-                                              & Robocasting                     & DOI: 10.1007/s00170-015-7576-2     \\
Direct Metal Deposition                        & -                               & DOI: 10.1007/s00170-015-7576-2     \\
Beam Interference Solidification               & -                               & DOI: 10.1007/s00170-015-7576-2     \\
Ballistic Particle Manufacturing               & -                               & DOI: 10.1007/s00170-015-7576-2     \\
Holographic Interference Solidification        & -                               & DOI: 10.1007/s00170-015-7576-2     \\
Solid Ground Curing                            & -                               & DOI: 10.1007/s00170-015-7576-2     \\
Solid Foil Polymerisation                      & -                               & DOI: 10.1007/s00170-015-7576-2     \\
Film Transfer Imaging                          & -                               & DOI: 10.1007/978-3-662-49866-8     \\
Cold Spraying  \textsuperscript{\textdied{}}                                 & Kaltgasspritzen \textsuperscript{\textdied{}}             & DOI: 10.1007/978-3-319-16772-5\_12 \\
Selective Deposition Lamination                & -                               & DOI: 10.1007/978-3-030-58960-8     \\
Digital Manufacturing \textsuperscript{\textdied{}}                        & Digitale Fertigung  \textsuperscript{\textdied{}}             & DOI: 10.1007/978-1-4939-2113-3     \\
-                                              & Direct Ink Writing              & DOI: 10.1002/adfm.200600434        \\
Stereolithography Apparatus                    & -                               & DOI: 10.1.1.564.3157               \\
    \midrule
    \textdied{ Keyword was eliminated} &       &  \\
    \label{table:Appendix_3d_printing_keywords}
  
\end{longtable}

\twocolumn

\clearpage
\pagebreak
\begin{table*}
\centering
\caption{Full names of the technical universities}
\label{tab:abbrevations_TUs}
\begin{tabular}{L{4cm} L{8cm}}
\textbf{Abbreviation} & \textbf{Full Name} \\
\hline
BTU & Brandenburgische Technische Universität Cottbus-Senftenberg \\
HCU & HafenCity Universität Hamburg \\
KIT & Karlsruher Institut für Technologie \\
LUH & Leibniz Universität Hannover \\
OVGU & Otto-von-Guericke-Universität Magdeburg \\
RWTH Aachen & Rheinisch-Westfälische Technische Hochschule Aachen \\
TU Berlin & Technische Universität Berlin \\
TU Braunschweig & Technische Universität Braunschweig \\
TU Chemnitz & Technische Universität Chemnitz \\
TU Darmstadt & Technische Universität Darmstadt \\
TU Dortmundt & Technische Universität Dortmund \\
TU Dresden & Technische Universität Dresden \\
TU Freiberg & Technische Universität Bergakademie Freiberg \\
TU Ilmenau & Technische Universität Ilmenau \\
TUC & Technische Universität Clausthal \\
TUHH & Technische Universität Hamburg \\
TUK & Technische Universität Kaiserslautern \\
TUM & Technische Universität München \\
U Stuttgart & Universität Stuttgart \\
\end{tabular}
\end{table*}

\pagebreak
\begin{landscape}
\begin{table}[htbp]
  \centering
  \caption{Industry classification}
    \begin{tabular}{lll}
    \textbf{Industry Name} & \multicolumn{1}{l}{\textbf{NACE range}} & \multicolumn{1}{l}{\textbf{Description}} \\
    \midrule
    Mining & \multicolumn{1}{l}{B} & \multicolumn{1}{l}{Mining and quarrying} \\
    Textiles/clothing & \multicolumn{1}{l}{C13-C15} & \multicolumn{1}{l}{Manufacturing of textiles, wearing apparel, fur, carpets, leather} \\
    Wood/paper/print & \multicolumn{1}{l}{C16-C18} & \makecell[l]{Manufacturing of wood and of products of wood and cork, except furniture; manufacture of articles of straw \\ and plaiting materials} \\
    Chemicals & \multicolumn{1}{l}{C20} & \multicolumn{1}{l}{Manufacturing of chemicals and chemical products} \\
    Pharmaceuticals & \multicolumn{1}{l}{C21} & \multicolumn{1}{l}{Manufacturing of basic pharmaceutical products and pharmaceutical preparations} \\
    Synthetics & \multicolumn{1}{l}{C22} & \multicolumn{1}{l}{Manufacturing of rubber and plastic products, packaging} \\
    Glass/ceramics & \multicolumn{1}{l}{C23} & \makecell[l]{Manufacturing of glass and glass products, refractory products,  clay building materials, tiles and flags, porcelain, \\ cement, concrete} \\
    Metal & \multicolumn{1}{l}{C24} & \multicolumn{1}{l}{Manufacturing of basic metals (iron, steel, ferro, lead, zinc and tin)} \\
    Metalware & \multicolumn{1}{l}{C25} & \multicolumn{1}{l}{Manufacturig of fabricated metal products (incl. cuttlery, tools), except machinery and equipment} \\
    Electronics/optics & \multicolumn{1}{l}{C26} & \multicolumn{1}{l}{Manufacturing of computer, electronic, optical products, and peripheral equipment} \\
    Electrical engineering & \multicolumn{1}{l}{C27} & \multicolumn{1}{l}{Manufacturing of electrical equipment (incl. electric motors, machinery \& equipment, appliances)} \\
    Mechanical engineering  & \multicolumn{1}{l}{C28} & \multicolumn{1}{l}{Manufacturing of machinery and equipment (incl. tools)} \\
    Automotive manufacturing & \multicolumn{1}{l}{C29} & \multicolumn{1}{l}{Manufacturing of motor vehicles, trailers and transport equipment} \\
    Transportation manufacturing & \multicolumn{1}{l}{C30} & \multicolumn{1}{l}{Manufacturing of boats, trains, aircraft, spacecraft} \\
    Other manufacturing & \multicolumn{1}{l}{C31-C32} & \multicolumn{1}{l}{Manufacturing of furniture, coins, games \& toys, jewellery, medical and dental instruments and supplies} \\
    Repair/installation & \multicolumn{1}{l}{C33} & \multicolumn{1}{l}{Repair and installation of machinery and equipment} \\
    Construction & \multicolumn{1}{l}{F} & \multicolumn{1}{l}{Construction of buildings, roads, railways, bridges, civil engineering, demolition, drilling, roofing, other installations} \\
    Wholesale & \multicolumn{1}{l}{G46} & \makecell[l]{Wholesale on a fee or contract basis (including raw materials, live animals, textile raw materials, household goods, \\ pharmaceuticals, equipment and supplies)} \\
    Retail & \multicolumn{1}{l}{G47} & \multicolumn{1}{l}{Retail trade of food, fuel, ICT, electronics, cultural goods, media, toys, flowers, plants} \\
    Media/publishing & \multicolumn{1}{l}{J58-J60} & \makecell[l]{Publishing of books, periodicals and other publishing activities,  motion picture, video and \\ television programme production, sound recording and music publishing activities, radio broadcasting} \\
    ICT services & \multicolumn{1}{l}{J61-J63} & \makecell[l]{Telecommunications, computer programming,  information service activities, data processing, hosting and  \\ related activities; web portals} \\
    Finance/insurance & \multicolumn{1}{l}{K} & \multicolumn{1}{l}{Financial and insurance activities, monetary intermediation, activities of holding companies, } \\
    Management services & \multicolumn{1}{l}{M69-M70} & \multicolumn{1}{l}{Legal and accounting activities, including  accounting, bookkeeping and auditing activities; tax consultancy} \\
    Engineering/scientific services & \multicolumn{1}{l}{M71-M72} & \multicolumn{1}{l}{Architectural and engineering activities; technical testing and analysis, scientific R\&D} \\
    Creative Services & \multicolumn{1}{l}{M73 } & \multicolumn{1}{l}{Advertising and market research, media representation, public opinion polling} \\
    Other services & \multicolumn{1}{l}{M74-N82} & \makecell[l]{Other professional, scientific and technical activities; Specialised design activities; Veterinary activities; Administrative \\ and support service activities; Rental and leasing activities;  Activities of employment placement agencies} \\
    Public administration & \multicolumn{1}{l}{O} & \multicolumn{1}{l}{Public administration and defence; compulsory social security; foreign affairs, defence, fire service} \\
    Education & \multicolumn{1}{l}{P} & \multicolumn{1}{l}{Education, driving schools, higher education, educational support activities} \\
    Interest groups & \multicolumn{1}{l}{S94} & \multicolumn{1}{l}{Activities of membership organisations, business associations, trade unions, political and religious organisations} \\
    Personal services & \multicolumn{1}{l}{S95-S96 and T} & \multicolumn{1}{l}{Repair of computers and personal and household goods, repair of computers and communication equipment} \\
    \midrule
    Source: https://nacev2.com/en &       &  \\
    \end{tabular}%
  \label{tab:sector_classification}
\end{table}%
 \end{landscape}   

\end{document}